\documentclass[twocolumn,showpacs,preprintnumbers,amsmath,amssymb,prb]{revtex4}
\usepackage{amsmath,amsfonts,bm,graphicx}
\usepackage{color}

\begin{document}

\title{Optical Hall conductivity in bulk and nanostructured graphene beyond the Dirac approximation}

\author{
Jesper Goor Pedersen$^1$, 
Mikkel H. Brynildsen$^2$,
Horia D. Cornean$^2$,
and Thomas Garm Pedersen$^{1,3}$
}
\affiliation{
$^1$
Department of Physics and Nanotechnology,
Aalborg University, Skjernvej 4A
DK-9220 Aalborg East, Denmark
\\
$^2$
Department of Mathematical Sciences, Aalborg University, Frederik Bajers Vej 7G, 9220 Aalborg East, Denmark
\\
$^3$
Center for Nanostructured Graphene (CNG), Aalborg University, DK-9220 Aalborg East, Denmark
}
\date{\today}

\begin{abstract}
We present a perturbative method for calculating the optical Hall conductivity in a tight-binding
framework based on the Kubo formalism. The method involves diagonalization only of the Hamiltonian in absence of 
the magnetic field, and thus avoids the computational problems usually arising due to the huge magnetic unit cells required
to maintain translational invariance in presence of a Peierls phase. 
A recipe for applying the method to numerical calculations of the magneto-optical response is presented.
We apply the formalism to the case of ordinary and gapped graphene in a next-nearest neighbour tight-binding model as
well as graphene antidot lattices. In both case, we find unique signatures in the Hall response, that are not 
captured in continuum (Dirac) approximations. These include a non-zero optical Hall conductivity
even when the chemical potential is at the Dirac point energy. 
Numerical results suggest that this effect should be measurable in experiments.
\end{abstract}

\pacs{81.05.ue, 78.20.Ls,78.20.Bh}

\maketitle

\section{Introduction}
Since the experimental discovery of graphene,\cite{Novoselov2004} the honeycomb lattice has been
the subject of intense research.\cite{Geim2007,Geim2009,Neto2009}
Graphene displays unique properties in an external magnetic field, with a non-equidistant Landau level structure and a zeroth 
Landau level energy which is independent of the magnetic field strength.\cite{Novoselov2005,Gusynin2005,Gusynin2005a}
The Landau level structure is reflected in the half-integer quantum Hall effect observed in graphene,\cite{Zhang2005} which, due to graphene's 
large cyclotron gap, has been observed at room temperature.\cite{Novoselov2007} These and other remarkable features 
of graphene emerge quite naturally from a low-energy, continuum description of graphene, the so-called Dirac approximation, 
which is based on a linearization of a nearest-neighbor (NN) tight-binding (TB) model near the high-symmetry $K$-points.\cite{Semenoff1984}

While many properties of graphene are correctly described by the Dirac model, it nevertheless fails in certain respects. A simple example 
is found in the optical response displaying a clear saddle point resonance around 4.4 eV that is absent in a linearized 
model.\cite{Pedersen2003a,Kravets2010}
More subtle effects such as an orbital magnetic susceptibility away from the Dirac point have been identified as consequences of lattice
effects lost in a continuum approach.\cite{Gomez-Santos2011} Here, we demonstrate that under certain circumstances the Hall conductivity, 
which is routinely applied as an important tool to characterize graphene experimentally,\cite{Kim2009,Bolotin2008,Zhang2005}
cannot be accurately described by the Dirac model. Specifically, the perfect electron-hole symmetry of the Dirac model, present also in the non-linearized NN TB model, results in an 
optical Hall conductivity which is identically zero unless electron-hole symmetry is broken by moving the chemical potential away 
from the Dirac point energy.\cite{Pedersen2003, Pedersen2011} In the present work, we predict that going beyond the Dirac model by including next-nearest neighbor coupling in a full TB model yields an appreciable optical Hall response even when the chemical potential coincides with the Dirac point.
This result is demonstrated using a novel perturbative technique that allows us to evaluate the magneto-optical response in an atomistic model for arbitrarily small field strengths. Thus, we identify a significant lattice effect, for which the Dirac model predicts a null result. We note
that, interestingly, this deviation occurs for energies well within the range of the linearized band structure of graphene. It is thus not a 
result of simply probing the band structure beyond the validity of the linearized model, as is the saddle point resonance mentioned above, 
but is rather a strong signature of broken electron-hole symmetry.

To go beyond the Dirac model, we return to the tight-binding model from which the Dirac approximation emerges.
The effect of a magnetic field can then be included via a Peierls substitution. The trouble with this method is that the periodicity
of the Peierls phase, for realistic magnetic field strengths, is usually orders of magnitude larger than the lattice constant.
Thus, calculations must be made on magnetic unit cells hundreds or thousands of times larger than the Wigner--Seitz cell.
In the case of bulk materials, where the unit cell consists of just a few atoms, this problem may
be overcome for large, but reasonably realistic magnitudes of the magnetic field. 
However, for nanostructured graphene materials, where the Wigner--Seitz cell may itself contain 
hundreds of atoms, direct diagonalization of the resulting Hamiltonian is not feasible. Several numerical methods
have previously been suggested to overcome this problem.\cite{Haydock1972,Czycholl1988,Oakeshott1993}
However, all these methods eventually fail at sufficiently small magnetic fields, because of the divergence of the 
size of the magnetic unit cell.

In this paper, we present a perturbative approach to calculating the
optical Hall conductivity in TB models on a honeycomb lattice. The approach requires diagonalization
only of the Hamiltonian in \emph{absence} of the magnetic field, and thus circumvents the problem associated with
the periodicity of the Peierls phase. We apply the formalism first to ordinary and gapped graphene, illustrating clear and
qualitative deviations from a Dirac approximation, as discussed above.
Finally, to illustrate the power of the perturbative formalism, we apply the method to an example of nanostructured graphene, 
in this case graphene antidot lattices,\cite{Pedersen2008,Furst2009}
and once again find qualitative differences in the optical Hall conductivity compared to a simple continuum treatment.

\section{Perturbative method}
The derivation of the main result of the perturbative treatment is given in full details in
Ref.~\onlinecite{Brynildsen2012}. 
The approach is based on the strategy developed in Refs.~\onlinecite{Cornean2006} and \onlinecite{Cornean2009}.
The starting point is a TB approximation
of the honeycomb lattice without magnetic field. The optical conductivity is then evaluated using the
Kubo formalism, with the effect of the magnetic field included perturbatively via a Peierls
substitution. That is, a phase is added to the hopping terms $t_{ij}$ between atomic sites $i$ and $j$, such that
$t_{ij}\rightarrow t_{ij} e^{i\phi}$, with the phase given as
$\phi=e/\hbar \times \int_\mathbf{R_i}^{\mathbf{R_j}}\mathbf{A}\cdot d\mathbf{l}$. Here, $\mathbf{R_i}$ and
$\mathbf{R_j}$ denote the positions of the atomic sites, while $\mathbf{A}$ is the magnetic vector potential.
Including the effect of the magnetic field to first order, the result for
the optical Hall conductivity reads as
\begin{eqnarray}
\sigma_{xy}(\omega) &=& 
\frac{B e^3}{16\pi^3\hbar^3\omega}\int\!\!d\mathbf{k} \mathrm{Re}\oint_\mathcal{C}\!\!dz\; 
\left\{i f(z)\right.
\nonumber \\
&&\times \left.\mathrm{Tr}\left[T_{xy}(z)+T_{xy}(z+\hbar\omega)\right]\right\}.\label{eq:sxy}
\end{eqnarray}
Also, we note that all even powers of the expansion in the magnetic field strength are zero, so the equation is
valid up to third order in the magnetic field strength.
Here, $B$ is the magnetic field strength, $e$ the electron charge, $\hbar\omega$ is the photon energy
and $f(z)$ is the Fermi-Dirac distribution function. 
The first integral is over the Brillouin zone, while the last
integral is to be taken along a contour $\mathcal{C}$, 
which should enclose the entire
energy spectrum of the Hamiltonian, $H$, while $\mathcal{C}\pm\hbar\omega$ should not contain the spectrum.
Here, and in what follows, we include a small imaginary part in $\hbar\omega=\hbar\omega_0+i\hbar\Gamma$ to
account for broadening.
Note that a spin degeneracy factor has not been included, and the final trace thus includes
tracing over spin degrees of freedom.
This trace is over the operators $T_{xy}(z)\equiv T_{xy}^{(1)}(z)+T_{xy}^{(2)}(z)$, with
\begin{eqnarray}
T_{xy}^{(1)}(z) 
&\equiv&
\left\{
\tilde{G}_yH_yG_x-\tilde{G}_xH_yG_y+
\right.\nonumber\\
&&
\left.
\tilde{G}\Big[H_yG\left(H_yG_x-H_xG_y\right)
\right.\nonumber \\
&&\left.
+(H_y\tilde{G}_x-H_x\tilde{G}_y)H_yG\Big]
\right\}H_x,
\end{eqnarray}
and
\begin{eqnarray}
T_{xy}^{(2)}(z) 
&\equiv&
\left[
\left(\tilde{G}_yH_{xy}-\tilde{G}_xH_{yy}\right)G
\right.\nonumber \\
&&\left.+
\tilde{G}\Big(H_{yy}G_x-H_{xy}G_y\Big)
\right]H_x.
\end{eqnarray}
Here, we have defined the derivatives of the Hamiltonian 
$H_i\equiv \partial H/\partial k_i$ and $H_{ij}\equiv \partial^2H/\partial k_i\partial k_j$, with
similar definitions for the Green's functions $G=(H-z)^{-1}$ and $\tilde{G}=(H-z+\hbar\omega)^{-1}$.
We stress that because the magnetic field is treated as a perturbation, the Hamiltonian appearing in
these expressions is the Hamiltonian in \emph{absence} of the magnetic field. For numerical calculations, this is 
a significant advantage of this method, as we will demonstrate in more detail below.

\begin{figure}
\begin{center}
\includegraphics[width=\linewidth]{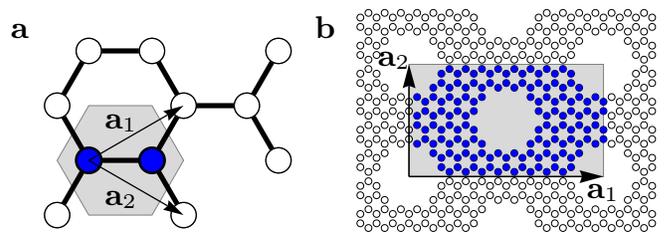}
\caption{(Color online) 
Unit cells used for the analytical and numerical calculations of the optical Hall conductivity
of (a) bluk graphene, and (b) a $\{4,2\}$ graphene antidot lattice.
The grey shading indicates the size of the unit cell, with highlighted carbon atoms included in the unit cell.
Note that the rectangular unit cell for the GAL is chosen for computational convenience only.
}
\label{fig:unitcell}
\end{center}
\end{figure}

\section{Linearized graphene}
Before turning to the full TB model, we wish to apply the above approach to demonstrate that, indeed, the Hall conductivity in the linearized model vanishes if electron-hole symmetry remains unbroken. We consider the TB Hamiltonian corresponding to 
the unit cell of graphene shown in Fig.~\ref{fig:unitcell}a. For generality, in addition to ordinary bulk graphene, we consider also
a gapped graphene model, where a staggered mass term $\pm \Delta$ is added to the on-site energies, with the sign alternating between the 
two sublattices of graphene. Pristine graphene may exhibit a gap due to excitonic effects.\cite{Khveshchenko2001,Gusynin2006} However,
here we will focus on magnitudes of the gap that are relevant for nanoengineered graphene, such as, e.g., graphene antidot
lattices.\cite{Pedersen2008,Furst2009} We stress that the results obtained remain qualitatively the same for any magnitude of the gap.
Linearized around the $K$-point, we find that the optical Hall conductivity can be evaluated as
\begin{equation}
\frac{\sigma_{xy}(\omega)}{\sigma_0} = \frac{4\omega_c^2}{\pi\omega^2}\int_\Delta^\infty\!\!\!d\epsilon\;
\frac{\hbar^2\omega^2+2\Delta^2-2\epsilon^2}{\hbar^2\omega^2-4\epsilon^2}
\left[f^\prime(-\epsilon)-f^\prime(\epsilon)\right],\label{eq:linsxy}
\end{equation}
where we have introduced the zero-frequency graphene conductivity $\sigma_0\equiv e^2/4\hbar$ and
the cyclotron frequency $\omega_c\equiv v_F\sqrt{2eB/\hbar}$, with $v_F$ the Fermi velocity. We note that the $K$ and $K^\prime$ valleys
contribute equally, and that a factor of 2 to account for this valley degeneracy has already been included
in the above equation.
This result agrees with the low field strength limit of previous analytical results derived by
Gusynin \emph{et al.}.\cite{Gusynin2007}
The full details of the derivation of 
Eq.~(\ref{eq:linsxy}) are given in the appendix. We note that the final term 
$f^\prime(-\epsilon)-f^\prime(\epsilon)\propto \sinh(\mu/kT)$ with $\mu$ the chemical potential while $kT$ is
the thermal energy.
This demonstrates how the optical Hall conductivity is identically zero in the symmetrical case, where the chemical 
potential sits at the Dirac point energy. Below, we will demonstrate how this zero-result is drastically altered when going
beyond the continuum (Dirac) treatment of graphene.

\section{Numerical results}
The analytical result for linearized graphene, presented above, is interesting in its own right, 
and serves as a way of 
validating the perturbative approach. However, the real power of the method lies in the fact
that because the expression in Eq.~(\ref{eq:sxy}) is given
in terms of the Hamiltonian \emph{without} magnetic field, numerical TB calculations can be performed on a 
drastically smaller unit cell than using the standard method of Peierls substitution in a non-perturbative manner.
Peierls substitution necessitates a magnetic unit cell large enough to ensure periodicity of the magnetic phase factor
added to the hopping terms. For graphene, this leads to a scaling of the total number of carbon atoms as $N\simeq 316\cdot10^3~\mathrm{T}\times B^{-1}$, rendering realistic magnetic fields quite difficult to manage using this method.\cite{Pedersen2011}

\subsection{Numerical recipe}
To arrive at an expression suitable for numerical simulations, we first note the identity $G_i=-GH_i G$,
with $H_i\equiv \partial H/\partial k_i$, and a similar definition for the Green's functions $G=(H-z)^{-1}$
and $\tilde{G}=(H-z+\hbar\omega)^{-1}$.
Using this identity we write the trace in the eigenstate basis as
\begin{eqnarray}
\mathrm{Tr}\left[T_{xy}^{(1)}(z)\right]
&=& 
\sum_{mnpq}\left\{
\frac
{M^{xxyy}_{mnpq}-M^{yxxy}_{mnpq}}
{\left(E_m\!-\!z_-\right)\left(E_n\!-\!z_-\right)\left(E_p\!-\!z\right)\left(E_q\!-\!z\right)}
\right.\nonumber \\
&+&\left.
\frac
{M^{xyxy}_{mnpq}-M^{xxyy}_{mnpq}}
{\left(E_m\!-\!z_-\right)\left(E_n\!-\!z\right)\left(E_p\!-\!z\right)\left(E_q\!-\!z\right)}
\right.\nonumber \\
&+&\left.
\frac
{M^{yxxy}_{mnpq}-M^{xyxy}_{mnpq}}
{\left(E_m\!-\!z_-\right)\left(E_n\!-\!z_-\right)\left(E_p\!-\!z_-\right)\left(E_q\!-\!z\right)}\right\},
\label{eq:trace1}
\end{eqnarray}
where we have introduced $z_\pm=z\pm\hbar\omega$ and 
\begin{equation}
M^{ijkl}_{mnpq} = \left<m|H_i|n\right> \left<n|H_j|p\right> \left<p|H_k|q\right> \left<q|H_l|m\right>.
\end{equation}
We can now perform the contour integration using the residue theorem.
To ease notation we define $E_{mn}=E_m-E_n$,
$M^{ijkl,i^\prime j^\prime k^\prime l^\prime}_{mnpq}=M^{ijkl}_{mnpq}-M^{i^\prime j^\prime k^\prime l^\prime}_{mnpq}$
as well as $\bar{\delta}_{mn}=1-\delta_{mn}$
and $\bar{\delta}_{mnp}=\bar{\delta}_{mn}\bar{\delta}_{np}\bar{\delta}_{pm}$, where $\delta_{mn}$ is
the Kronecker delta. 
We then arrive at the rather lengthy expression
\begin{widetext}
\begin{eqnarray}
\oint_\mathcal{C} f(z)\mathrm{Tr}\left[T_{xy}^{(1)}(z)\right]dz
&=&
2\pi i\sum_{mnpq}\left\{
\bar{\delta}_{pq}
\frac{M^{xxyy,yxxy}_{mnpq}}{E_{pq}}
\left(
\frac{f(E_p)}{(E_{mp}+\Omega)(E_{np}+\Omega)}
-
\frac{f(E_q)}{(E_{mq}+\Omega)(E_{nq}+\Omega)}
\right)
\right.\nonumber \\
&&\left.+
\delta_{pq}
M^{xxyy,yxxy}_{mnpp}
\frac
{\left(E_{mp}+E_{np}+2\Omega\right)f(E_p)+\left(E_{mp}+\Omega\right)\left(E_{np}+\Omega\right)f^\prime(E_p)}
{\left(E_{mp}+\Omega\right)^2\left(E_{np}+\Omega\right)^2}
\right.\nonumber \\
&&\left.+
\bar{\delta}_{npq}\left(M^{xyxy,xxyy}_{mnpq}+M^{xyxy,xxyy}_{mpnq}+M^{xyxy,xxyy}_{mpqn}\right)
\frac{f(E_n)}{E_{np}E_{nq}\left(E_{nm}-\Omega\right)}
\right.\nonumber \\
&&+\left.
\delta_{np}\bar{\delta}_{nq}\frac{M^{xyxy,xxyy}_{mnnq}+M^{xyxy,xxyy}_{mnqn}+M^{xyxy,xxyy}_{mqnn}}{E_{nq}^2}
\right.\nonumber \\
&&\times\left.
\left(
\frac
{\left(E_{mn}+E_{qn}+\Omega\right)f(E_n)+E_{qn}\left(E_{mn}+\Omega\right)f^\prime(E_n)}
{\left(E_{mn}+\Omega\right)^2}+\frac{f(E_q)}{E_{qm}-\Omega}
\right)
\right.\nonumber \\
&&-\left.
\delta_{nq}\delta_{pq}M^{xyxy,xxyy}_{mnnn}
\frac
{\left(E_{mn}+\Omega\right)f^\prime(E_n)+\frac{1}{2}\left(E_{mn}+\Omega\right)^2f^{\prime\prime}(E_n)+f(E_n)}
{\left(E_{mn}+\Omega\right)^3}
\right.\nonumber \\
&&-\left.
\frac{M^{yxxy,xyxy}_{mnpq}f(E_q)}{\left(E_{mq}+\Omega\right)\left(E_{nq}+\Omega\right)\left(E_{pq}+\Omega\right)}
\right\},\label{eq:contour1}
\end{eqnarray}
\end{widetext}
where we have introduced $\Omega=\hbar\omega$.
In a similar fashion, we write the trace over the second operator as
\begin{eqnarray}
\mathrm{Tr}\left[T_{xy}^{(2)}(z)\right]
&=&
\sum_{mnp}\left\{
\frac
{N^{yxyx}_{pmn}-N^{yxxy}_{pmn}}
{\left(E_m-z_-\right)\left(E_n-z\right)\left(E_p-z\right)}
\right.\nonumber\\
&&\left.
-\frac
{N^{xxyy}_{mnp}-N^{yxxy}_{mnp}}
{\left(E_m-z_-\right)\left(E_n-z_-\right)\left(E_p-z\right)}
\right\},\label{eq:trace2}
\end{eqnarray}
where we have defined
\begin{equation}
N^{ijkl}_{mnp} = \left<m|H_i|n\right> \left<n|H_{jk}|p\right> \left<p|H_l|m\right>.
\end{equation}
The residue theorem then leads to
\begin{eqnarray}
&&\oint_\mathcal{C} f(z)\mathrm{Tr}\left[T_{xy}^{(2)}(z)\right]dz
= \nonumber \\
&&
2\pi i\sum_{mnp}\left\{
\left(N_{mnp}^{xxyy,yxxy}\right)\frac{f(E_p)}{\left(E_{mp}+\Omega\right)\left(E_{np}+\Omega\right)}
\right.\nonumber \\
&&\left.
+\bar{\delta}_{np}\frac{N_{pmn}^{yxyx,yxxy}}{E_{np}}
\left(
\frac{f(E_n)}{E_{mn}+\Omega}-\frac{f(E_p)}{E_{mp}+\Omega}
\right)
\right.\nonumber \\
&&\left.
+\delta_{np}N_{nmn}^{yxyx,yxxy}
\frac{f(E_n)+\left(E_{mn}+\Omega\right)f^\prime(E_n)}{\left(E_{mn}+\Omega\right)^2}
\right\}.\label{eq:contour2}
\end{eqnarray}
The trace of the operators with the shifted argument is obtained in a similar fashion. By substituting 
$z\rightarrow z_+$ and $z_-\rightarrow z$ in Eq.~(\ref{eq:trace1}) and relabeling slightly, one can show
that the contour integral $\oint_\mathcal{C} f(z)\mathrm{Tr}[T^{(1)}_{xy}(z+\Omega)]$ is given 
by Eq.~(\ref{eq:contour1}), if one substitutes
\begin{eqnarray}
\Omega &\rightarrow& -\Omega \nonumber \\
M_{mnpq}^{xxyy,yxxy} &\rightarrow& M_{qpnm}^{xxyy,yxxy} \nonumber \\
M_{mnpq}^{yxxy,xyxy} &\rightarrow& M_{qpnm}^{xyxy,xxyy} \nonumber \\
M_{mnpq}^{xyxy,xxyy} &\rightarrow& M_{qpnm}^{yxxy,xyxy}.
\end{eqnarray}
Similarly, $\oint_\mathcal{C} f(z)\mathrm{Tr}[T^{(2)}_{xy}(z+\Omega)]$ is given Eq.~(\ref{eq:contour2}),
with the substitutions
\begin{eqnarray}
\Omega &\rightarrow& -\Omega \nonumber \\
N_{nmp}^{xxyy,yxxy} &\rightarrow& N_{mpn}^{yxxy,yxyx} \nonumber \\
N_{pnm}^{yxyx,yxxy} &\rightarrow& N_{pnm}^{yxxy,xxyy}.
\end{eqnarray}
In this way we have arrived at expressions for the contour integral of the trace in Eq.~(1) 
in terms of sums over the eigenstates of the Hamiltonian without magnetic field. These sums can be truncated
to include only states near the Fermi energy. For small magnetic fields, this method is drastically faster than
direct diagonalization of the Hamiltonian with magnetic field, the size of which diverges as the magnetic field
strength is reduced.
In all numerical results presented below, we set the thermal energy to $kT=0.025$~eV and include a 
broadening of $\hbar\Gamma=0.05$~eV.

\subsection{Graphene}
We now consider a next-nearest neighbor (NNN) TB model of gapped graphene,
defined via the Hamiltonian
\begin{equation}
H(\mathbf{k}) = \left[
\begin{array}{cc}
t^\prime g(\mathbf{k})+\Delta & -t f(\mathbf{k}) \\
-t f^*(\mathbf{k}) & t^\prime g(\mathbf{k})-\Delta
\end{array}
\right],\label{eq:HNNN}
\end{equation}
parametrized by the nearest and next-nearest neighbor hopping
parameters $t=3$~eV and $t^\prime = 0.3$~eV, respectively. 
Here, 
$f(\mathbf{k})=e^{ik_xa_c}+2e^{-ik_xa_c/2}\cos(\sqrt{3}k_ya_c/2)$,
while
$g(\mathbf{k})=2\cos(\sqrt{3}k_y a_c)+4\cos(3k_x a_c/2)\cos(\sqrt{3}k_ya_c/2)$, where $a_c=a/\sqrt{3}$ is the
carbon-carbon distance.
While the hopping parameters can vary slightly depending on which ab initio results they are fitted to, we note that the
exact value of the hopping terms do not alter out results qualitatively. 
We set the on-site energy to zero.
For $t^\prime=0$, this  model has electron-hole symmetry, which is inherited in the Dirac model discussed above. 
We note that linearization of any TB model will inevitably result in perfect electron-hole symmetry.
As discussed above, in the fully symmetrical situation, where the chemical potential sits at the Dirac point energy,
the off-diagonal conductivity is identically zero for any such model.
This can be proven on quite general terms for all TB models exhibiting $\pi$--$\pi^*$ symmetry, for which contributions
from conjugate transitions exactly cancel in the fully symmetrical case.\cite{Pedersen2003, Pedersen2011}
Introducing next-nearest neighbor coupling breaks electron-hole symmetry and, 
as we will now demonstrate, leads to a markedly different magneto-optical response of graphene.

\begin{figure}
\begin{center}
\includegraphics[width=\linewidth]{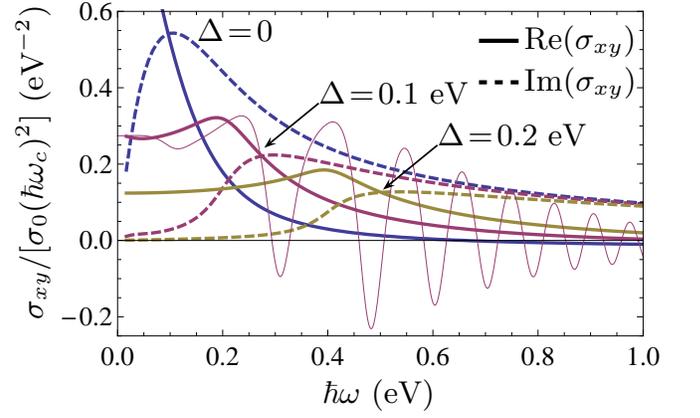}
\caption{(Color online) Optical Hall conductivity, $\sigma_{xy}$, as a function of photon energy for
ordinary graphene and gapped graphene with increasing values of the mass term. The chemical potential is in the middle of the gap.
The conductivity is shown relative to the zero-frequency conductivity of graphene, $\sigma_0$, times the
square of the cyclotron energy, $\hbar\omega_c$. The solid (dashed) lines show the real (imaginary) part of the
conductivity. To ease visibility, the response of ordinary graphene has been cropped. The DC value of ordinary graphene
is $\sigma_{xy}(0) \simeq (\hbar\omega_c)^2\sigma_0\times 1.03$~eV$^{-2}$.
The thin line shows the real part of the Hall conductivity for $\Delta=0.1$~eV and a magnetic field of $B=26.3$~T.
}
\label{fig:sxymu0}
\end{center}
\end{figure}
In Fig.~\ref{fig:sxymu0} we show the calculated optical Hall conductivity of graphene, with the chemical potential
at the Dirac point energy. While the Dirac approximation (and nearest-neighbor TB) predicts a zero response in this case, our NNN TB model 
suggests a clear resonance at $\hbar\omega=2\Delta$.
This drastic deviation from the Dirac model is due to the broken electron-hole symmetry, which
means that conjugate transitions in the optical response no longer cancel entirely.\cite{Pedersen2011}
The strength of the resonance decreases as the magnitude of the band gap 
is increased, in agreement with previous results showing that a sufficiently large band gap effectively quenches
the effect of the magnetic field, provided $\Delta\gg \hbar\omega_c$.\cite{Pedersen2011}
However, the magnitude of this correction to the Dirac response is quite appreciable, suggesting that these deviations 
from the Dirac model should be measurable in experiments.
We note that, as expected, numerical calculations show similar results for a nearest-neighbor model if overlap between 
neighboring $\pi$ orbitals is included.

For comparison with the perturbative results, we also show the Hall conductivity for $\Delta=0.1$~eV and a magnetic field
strength of $B=26.3$~T, calculated using standard, non-perturbative tight-binding methods.\cite{Pedersen2011} We note
that these calculations involve a Hamiltonian with $12000 \times 12000$ elements for what is a quite strong
magnetic field. The relationship with the perturbative result is evident in the figure, and illustrates the fact that
the perturbative results still have predictive power for the strength of the response even in substantial magnetic fields.
In particular, the perturbative results correspond to an averaging of the oscillations occurring due to individual Landau levels,
which for smaller magnetic field strengths could presumably be caused by a broadening of the order of the cyclotron energy.

\begin{figure}
\begin{center}
\includegraphics[width=\linewidth]{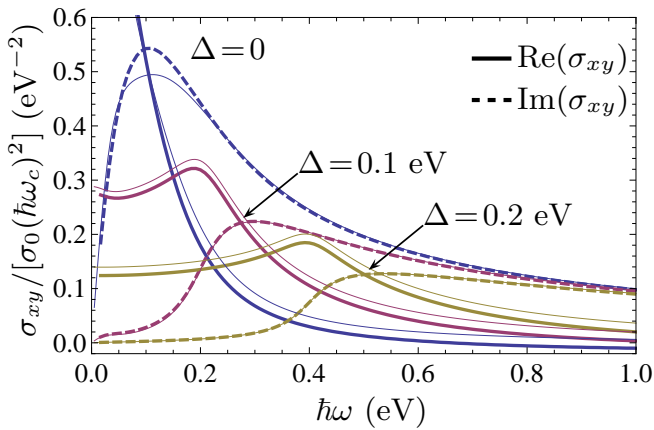}
\caption{(Color online) Optical Hall conductivity, $\sigma_{xy}$, as a function of photon energy for
ordinary graphene and gapped graphene with increasing values of the mass term. The chemical potential is in the middle of the gap.
The conductivity is shown relative to the zero-frequency conductivity of graphene, $\sigma_0$, times the
square of the cyclotron energy, $\hbar\omega_c$. The solid (dashed) lines show the real (imaginary) part of the
conductivity. The thin lines show the corresponding results obtained via the semi-analytical expression derived
in the main text.
}
\label{fig:analNNN}
\end{center}
\end{figure}
To further corroborate these findings, we will now derive an approximate, semi-analytical expression for the optical Hall conductivity
in the next-nearest neighbor model. We first note that linearizing the NNN Hamiltonian in Eq.~(\ref{eq:HNNN}) near the $K$ point
yields the same result as the nearest-neighbor model, except for a constant diagonal term. Instead, we proceed by expanding the diagonal
NNN term to second order near the $K$ point, yielding the approximate Hamiltonian
\begin{equation}
H(\mathbf{k}) \simeq \left[
\begin{array}{cc}
\tfrac{9}{8}\tau\kappa^2+\Delta & \tfrac{3}{2}(\kappa_x-i\kappa_y) \\
\tfrac{3}{2}(\kappa_x+i\kappa_y) & \tfrac{9}{8}\tau\kappa^2-\Delta
\end{array}
\right],
\end{equation}
where we have defined $\kappa_i=tk_i a_c$ and introduced the parameter $\tau=2t^\prime/t^2$, quantifying the perturbation
due to NNN coupling. The eigenvalues of this Hamiltonian read $E_\pm=\tfrac{9}{8}\tau\kappa^2\pm\sqrt{\Delta^2+\tfrac{9}{4}\kappa^2}$.
We now proceed in a manner similar to that of the appendix, i.e., we evaluate the trace of $T_{xy}(z)$, integrate out
the angular component of the Brillouin zone integral and then use the residue theorem to perform the contour integral over $z$ in
Eq.~(\ref{eq:sxy}). In this manner, we find that the optical Hall conductivity in the NNN model is approximately given by
\begin{widetext}
\begin{equation}
\sigma_{xy}(\omega) =
\sigma_0\frac{4\omega_c^2}{\pi\omega^2}
\int_\Delta^\infty 
\frac{\tfrac{1}{2}\tau(3\epsilon^2-\Delta^2)\hbar^2\omega^2\left[f(E_+)-f(E_-)\right]+a_-(\epsilon)f^\prime(E_-)+a_+(\epsilon)f^\prime(E_+)}
{\epsilon^2(\hbar^2\omega^2-4\epsilon^2)}\;d\epsilon,
\end{equation}
where we have introduced
\begin{equation}
a_\pm(\epsilon) = \epsilon\left\{
2\epsilon(\tau \epsilon\pm 1)\left(1-\left(\tau \epsilon\right)^2\right)\left(\epsilon^2-\Delta^2\right)
-\hbar^2\omega^2
\left(
\pm \epsilon - \tfrac{1}{2}\tau\left[\left(\tau \epsilon\right)^2-1\pm2\tau \epsilon\right]\left(\epsilon^2-\Delta^2\right)
\right)
\right\},
\end{equation}
\end{widetext}
where $E_\pm = \tfrac{1}{2}\tau(\epsilon^2-\Delta^2)\pm\epsilon$, with $\epsilon=\sqrt{\Delta^2+\tfrac{9}{4}\kappa^2}$.
In Fig.~\ref{fig:analNNN} we compare the numerical results with those obtained by numerical integration of the 
analytical result derived above. We note that there is excellent agreement between the two methods.

\subsection{Graphene antidot lattices}
Nanostructured graphene systems, with Wigner-Seitz unit cells containing on the order of hundreds of atoms, are practically impossible 
to handle using direct diagonalization of the TB Hamiltonian in the presence of a realistic magnetic field. 
To illustrate the power of the perturbative method presented above,
we now consider the magneto-optical response of periodically perforated graphene, so-called
graphene antidot lattices (GALs).\cite{Pedersen2008}
The low-energy spectrum of these structures can be quite accurately described in a gapped graphene model, by fitting
the mass term to coincide with half the magnitude of the GAL band gap.\cite{Pedersen2009} We now compare TB results to such a continuum description
of GALs.
For these results we ignore next-nearest neighbor coupling, to illustrate how deviations from a continuum approximation emerge even
in the simplest nearest-neighbor TB treatment.
We denote a given GAL structure as $\{L, R\}$, where $L$ is the side length of the hexagonal
Wigner--Seitz cell, while $R$ denotes the radius of the circular hole in the middle of the cell, both in units of the graphene lattice constant.
We consider a geometry for which the superlattice basis vectors are parallel to the carbon-carbon bonds, as such structures always
exhibit band gaps.\cite{Petersen2011} As an example, Fig.~\ref{fig:unitcell}b shows the computational cell of a $\{4,2\}$ GAL,
highlighted with gray shading. We note that the rectangular unit cell contains 144 carbon atoms. For comparison, a standard non-perturbative 
calculation of the magneto-optical properties would require a magnetic unit cell consisting of 72000 carbon atoms, 
even for a substantial magnetic field strength of $40$~T.

\begin{figure}
\begin{center}
\includegraphics[width=\linewidth]{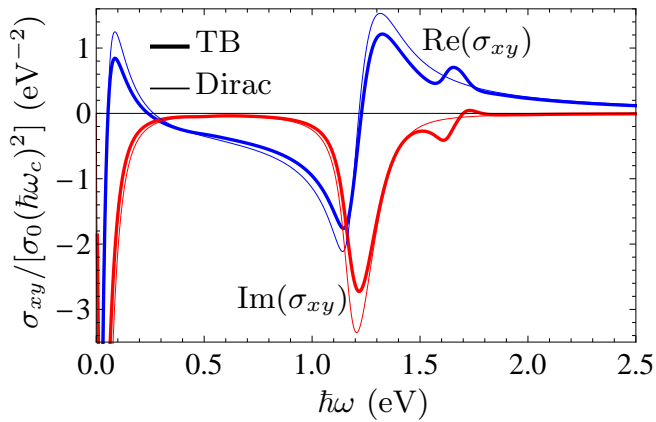}
\caption{(Color online) Optical Hall conductivity, $\sigma_{xy}$, as a function of photon energy for the
$\{4,2\}$ graphene antidot lattice.
The conductivity is shown relative to the zero-frequency conductivity of graphene, $\sigma_0$, times the
square of the cyclotron energy, $\hbar\omega_c$. The blue (red) lines show the real (imaginary) part of the
conductivity. The thick lines show the results of the perturbative method applied to the
GAL structure, while the thin lines are results of a gapped graphene Dirac model with a band gap corresponding
to the GAL.
}
\label{fig:sxyL4R2}
\end{center}
\end{figure}
In Fig.~\ref{fig:sxyL4R2} we show the optical Hall conductivity of the $\{4,2\}$ GAL calculated using the perturbative
approach. For comparison we also show the corresponding result for gapped graphene, with a mass term equal to 
half the band gap of the GAL, $\Delta\simeq 0.58$~eV. In both cases, we fix the chemical potential at the lower
band gap edge, i.e., $\mu=\Delta$. We find reasonable agreement between the full GAL calculations
and the simpler gapped graphene model. However, we note that a distinct feature of the GAL structure is the additional
resonance near $\hbar\omega=1.65$~eV, which is absent in the simpler gapped graphene Dirac model. This resonance occurs
due to transitions between bands that are not present in a simple two-band gapped graphene model of GALs.
We will explore the details of the magneto-optical response of graphene antidot lattices in future work, and include the
result here mainly to emphasize the power of the perturbative formalism presented.

\section{Summary}
A perturbative approach to calculating the optical Hall conductivity of graphene structures has been
presented and applied to tight-binding models of graphene and graphene antidot lattices. The optical
Hall response of graphene shows significant deviations from a simple Dirac treatment. While the Dirac model
predicts a Hall conductivity of identically zero for a chemical potential at the Dirac point energy, results from our next-nearest 
neighbor tight-binding model indicate clear resonances at the band gap. The numerical
results suggest that these effects should be measurable in experiments. Results for graphene antidot lattices
illustrate that in this case, even the simple nearest-neighbor tight-binding model gives qualitatively different
results than a simple Dirac approximation.

\section{Aknowledgments}
The work by J.G.P. is financially supported by the Danish Council for Independent Research, FTP grant numbers
11-105204 and 11-120941. The Center for Nanostructured Graphene (CNG) is sponsored by the Danish National 
Research Foundation.

\appendix
\section{Linearized graphene}
Linearizing the tight-binding Hamiltonian of gapped graphene near the $K$ point results in the celebrated
Dirac approximation of graphene,
\begin{equation}
H=\left[\begin{array}{cc}
\Delta & 3/2\left(\kappa_x-i\kappa_y\right) \\
3/2\left(\kappa_x+i\kappa_y\right) & -\Delta
\end{array}
\right],
\end{equation}
where  we have introduced $\kappa_i=tk_i a_c$, with $a_c$ the nearest-neighbor distance $a_c=a/\sqrt{3}$
We use this form of the Hamiltonian to evaluate the trace, noting that the linearization means that 
$T_{xy}^{(2)}=0$. In polar coordinates, $(\kappa_x,\kappa_y)=\kappa(\cos\phi, \sin\phi)$ we find,
after integrating over the angular component,
\begin{eqnarray}
&&\int d\phi\;\mathrm{Tr}\left[T_{xy}(z)\right] = \nonumber \\
&&\frac
{10368\pi t^4a_c^4\Omega(\Omega-2z)\left[z(\Omega-z)+\Delta^2 \right]}
{\left[9\kappa^2-4\left(\left(\Omega-z\right)^2-\Delta^2\right)\right]^2\Big[9\kappa^2-4\left(z^2-\Delta^2\right)\Big]^2},
\nonumber \\
\end{eqnarray}
where $\Omega=\hbar\omega$ is the photon energy.
We now use the residue theorem to perform the contour integral over $z$, yielding
\begin{eqnarray}
&&\oint_\mathcal{C}dz\; if(z) \int \!\!d\phi\;\mathrm{Tr}\left[T_{xy}(z)+ T_{xy}(z+\Omega)\right]=\nonumber \\ 
&&\frac
{81\pi^2t^4a_c^4\left(9\kappa^2-2\Omega^2\right)}
{2\Omega\sqrt{9\kappa^2+4\Delta^2}\left(9\kappa^2+4\Delta^2-\Omega^2\right)}
\nonumber \\
&&\times
\left[
f^\prime\left(-\sqrt{9\kappa^2/4+\Delta^2}\right)-f^\prime\left(\sqrt{9\kappa^2/4+\Delta^2}\right)
\right],
\end{eqnarray}
where poles at $z=2\Omega\pm\sqrt{9\kappa^2/4+\Delta^2}=2\Omega\pm  \epsilon(k)$ have been ignored because, 
as discussed in the paper, the contour explicitly excludes these points. 
Inserting this result in Eq.~(1) of the paper
and converting the Brillouin zone integration to an integral
over energy, this leads to Eq.~(\ref{eq:linsxy}) of the main text.
Taking as their starting point the Landau level structure 
of gapped graphene, Gusynin \emph{et al.} have previously derived the off-diagonal magneto-optical conductivity 
of gapped graphene.\cite{Gusynin2007} Their result is stated as a sum over Landau levels,
\begin{eqnarray}
\frac{\sigma_{xy}(\omega)}{\sigma_0} &=& \frac{2\hbar^2\omega_c^2}{\pi}\nonumber\\
&&\times 
\sum_{n=0}^\infty
\left(
\left[f(-\epsilon_{n+1})-f(-\epsilon_n)\right]-\left[f(\epsilon_{n})-f(\epsilon_{n+1})\right]\right)
\nonumber \\
&&
\times\left[
\left(1-\frac{\Delta^2}{\epsilon_n \epsilon_{n+1}}\right)
\frac{1}{(\epsilon_{n+1}-\epsilon_n)^2-\Omega^2}
\right.\nonumber\\
&&
\left.
+
\left(1+\frac{\Delta^2}{\epsilon_n \epsilon_{n+1}}\right)
\frac{1}{(\epsilon_{n+1}+\epsilon_n)^2-\Omega^2}
\right],\label{eq:sxyDE}
\end{eqnarray}
where the energies are $\epsilon_n = \sqrt{\Delta^2+n\hbar^2\omega_c^2}$ for $n \geq 0$,
yielding 
$\epsilon_{n+1}=\epsilon_n\sqrt{1+\hbar^2\omega_c^2/\epsilon_n^2}\simeq\epsilon_n+\hbar^2\omega_c^2/(2\epsilon_n)$
in the low-field limit. Thus, in the continuum limit $\frac{d\epsilon}{dn}=\hbar^2\omega_c^2/2$.
Replacing $f(\epsilon_{n})-f(\epsilon_{n+1})\rightarrow -f^\prime(\epsilon_n) \frac{d\epsilon}{dn}$ and converting
the sum to an integral via $\sum_n \frac{d\epsilon}{dn} \rightarrow \int d\epsilon$, we recover 
Eq.~(\ref{eq:linsxy}).


\end{document}